\documentclass[prd, onecolumn]{revtex4}

\usepackage{graphicx}
\usepackage{dcolumn}
\usepackage{bm}
\usepackage{wrapfig}
\usepackage{epsfig}
\usepackage{breqn}
\usepackage{amssymb}
\usepackage{subfigure}
 \usepackage{float}

\begin{document}


\title{Einstein's Field Equations as Continuous-Time Recurrent Neural Networks}

\author{Ikjyot Singh Kohli}
	\email{isk@mathstat.yorku.ca}
\affiliation{York University - Department of Mathematics and Statistics}

\date{December 29, 2019}

\begin{abstract}
We examine a subset of spatially homogenous and anisotropic solutions to Einstein's field equations: the Bianchi Type A models, and show that they can be written as a continuous-time recurrent neural network (CTRNN). This reformulation of Einstein's equations allows one to write potentially complicated nonlinear equations as a simpler dynamical system consisting of linear combinations of the neural network weights and logistic sigmoid activation functions. The CTRNN itself is trained by using an explicit Runge-Kutta solver to sample a number of solutions of Einstein's equations for the Bianchi Type A models and then using a nonlinear least-squares approach to find the optimal set of weights, time delay constants, and bias parameters that provide the best fit of the CTRNN equations to the Einstein equations. In terms of numerical examples, we specifically provide solutions to Bianchi Type I and II models. We conclude the paper with some comments on optimal parameter probability distributions and ideas for future work.
\end{abstract}
\maketitle 

\section{Introduction}
Artificial neural networks and deep learning have become an increasing part of everyday life because of their ability to learn deep patterns from data and then apply the subsequent learned structure to solve new and important problems. Part of this class of problems is determining whether neural networks can provide new insights into problems from physics. One of the main reasons for this is that fundamental laws of physics are usually written as tensor equations which when written in local coordinates take the form of systems of partial differential equations which can be difficult to solve analytically. Typical strategies to solve these equations involve exploiting symmetries, transforming coordinates, or numerical methods.  

Recently, Breen, Foley, Boekholt, and Zwart \cite{breen2019newton} used artificial neural networks to solve the Newtonian three-body problem much faster than typical state-of-the-art numerical solvers. Iten, Metger, Wilming, del Rio, and Renner \cite{iten2018discovering} derived a neural network architecture that can extract simple physical concepts from arbitrary experimental data. They were able to apply this neural network structure to gain insights from a variety of simple physical examples in classical and quantum mechanics from damped pendulums and two-particle collisions to qubits. Han, et.al \cite{Han8505} employed deep learning to solve parabolic partial differential equations such as the Allen-Cahn equation which is a reaction-diffusion type equation. Deep learning has also found applications in high energy physics with respect to mass reconstruction, jet substructure, and jet-flavour classification as discussed by Guest, Kranmer, and Whiteson \cite{doi:10.1146/annurev-nucl-101917-021019}.

In this paper, we apply the concept of continuous-time recurrent neural networks (CTRNNs) to Einstein's field equations to show that these equations can be written in a very simple form under suitable conditions which we discuss below in further detail. 


\section{Mathematical Necessities}
Our starting point is to write Einstein's field equations which are tensor equations as dynamical equations. We consider the $G_3$ group of cosmological models which consist of spatially homogeneous and anisotropic solutions of Einstein's field equations \cite{ellismac}. These models have the property that unit normals $\mathbf{n}$ to spatially homogeneous hypersurfaces are irrotational and geodesic and invariant under Killing vector transformations (the models are simply transitive). Hence, taking a basis of Killing vectors $\{ \xi_i \}$, we choose $t$ such that $\mathbf{n} = \partial_t$, and that $n_a = -\nabla_a t$. One can then choose a tetrad of vectors $\mathbf{e}_{a}$ with $\mathbf{n} = \mathbf{e}_{0}$ at a point $p$. One can then write a spacetime metric as \cite{elliscosmo}:
\begin{equation}
\label{eq:metric}
ds^2 = -dt^2  + \gamma_{ij} (e^i_\alpha dx^\alpha)(e^j_\beta dx^\beta),
\end{equation}
where $e^i_\alpha dx^\alpha$ are 1-forms inverse to the spatial vector triad $\mathbf{e}_i$, and $\gamma_{ij}$ denotes the spatial metric tensor.

We employ orthonormal frames, where $g_{ab} = \eta_{ab}$ and the time-dependence is then in the commutation functions for the basis vectors. Namely, we have that:
\begin{equation}
\left[\mathbf{e}_{a}, \mathbf{e}_{b}\right] = \gamma^c_{ab}(t) \mathbf{e}_{c}.
\end{equation}
Under this formalism, the Einstein field equations, Jacobi identities, and contracted Bianchi identities can be written in terms of what are known as basic variables:
\begin{equation}
\left(H, \sigma_{ab}, \dot{u}_{a}, \omega_a, \Omega_a, n_{ab}, a_a\right),
\end{equation}
and source terms
\begin{equation}
(\mu, p, q_a, \pi_{ab}).
\end{equation}
We refer the reader to \cite{ellismac} for the details of meaning of these variables. For example, $H$ is the Hubble parameter, $\sigma_{ab}$ denotes the shear tensor, $\omega_{a}$ denotes the vorticity vector, $n_{ab}$ and $a_{a}$ are curvature variables, and $\mu$ and $p$ are the matter energy density and pressure respectively.  

As mentioned in the introduction, in this work we consider only Bianchi Class A models. The physical state of such models is described by a vector $(H, \mathbf{x})$ where
\begin{equation}
\mathbf{x} = (\sigma_{+}, \sigma_{-}, n_1, n_2, n_3).
\end{equation}
From a dynamical systems perspective, it is useful to normalize these variables with respect to powers of the Hubble parameter $H$, thereby obtaining a reduced dynamical system. That is, the Einstein field equations become:
\begin{eqnarray}
\label{eq:evo1}
\Sigma_{\pm}' &=& -(2-q)\Sigma_{\pm} - S_{\pm}, \\
\label{eq:evo2}
N_{1}' &=& (q - 4\Sigma_{+})N_1, \\
\label{eq:evo3}
N_{2}' &=& (q + 2 \Sigma_{+} + 2\sqrt{3} \Sigma_{-})N_2, \\
\label{eq:evo4}
N_{3}' &=& (q +  2 \Sigma_{+} -2 \sqrt{3} \Sigma_{-}) N_3, 
\end{eqnarray}
where
\begin{eqnarray}
S_{+} &=& \frac{1}{6} \left[ (N_2 - N_3)^2 - N_1 (2N_1 - N_2 - N_3)\right], \\
S_{-} &=& \frac{1}{2 \sqrt{3}}(N_3 - N_2)(N_1 - N_2 - N_3),
\end{eqnarray}
and $'$ denotes differentiation with respect to a time variable $\tau$, defined by the relationship $dt/d\tau = H^{-1}$. Further, we have that
\begin{equation}
q = \frac{1}{2}(3 \gamma - 2)(1-K) + \frac{3}{2}(2- \gamma) \Sigma^2,
\end{equation}
where
\begin{equation}
K = \frac{1}{12} \left[N_1^2 + N_2^2 + N_3^2 - 2(N_1 N_2 + N_2 N_3 + N_3 N_1)\right],
\end{equation}
and $\gamma$ is an equation of state parameter that is used to classify the type of matter that is being used in the model, and for ordinary physical matter has values such that $0 \leq \gamma \leq 2$.

The physical state space of these models is described by an inequality derived from the Friedmann equation:
\begin{equation}
\label{eq:inequality1}
\Omega = 1 - \Sigma^2 - K = 1- \Sigma_{+}^2 - \Sigma_{-}^2 - K \geq 0.
\end{equation}

One sees that Eqs. (\ref{eq:evo1})-(\ref{eq:evo4}) are a complicated nonlinear system of equations. The goal of this paper is to show that these equations can be represented by a dynamical system of much simpler equations, which are continuous-time recurrent neural networks (CTRNNs). As shown by Funahashi and Nakamura \cite{funahashi}, CTRNNs are universal dynamics approximators, in the sense that for any finite interval of time, CTRNNs can approximate the trajectories of any smooth dynamical system on a compact subset of $\mathbb{R}^{n}$ arbitrarily well. Following \cite{doi:10.1177/105971239500300405},  we note that CTRNNs are defined as networks of model neurons of the following general form:
\begin{equation}
\label{eq:ctrnn1}
\dot{y}_{i} = f_{i} (y_1, \ldots, y_N) = \frac{1}{\tau_i} \left[-y_i + w^{j}_{i} \sigma(y_j + \theta_j) + I_i\right], \quad i = 1,2,\ldots,N.
\end{equation}
(Note that: we have used the Einstein summation convention where repeated upper and lower indices are summed.) In Eq. (\ref{eq:ctrnn1}), $y$ is the state of each neuron, $\tau$ is the associate time constant $(\tau > 0)$, $w_{ji}$ is the weight of the connection from the $j$th neuron to the $i$th nueron, $\theta$ is the bias term, $\sigma(x) = 1/(1+e^{-x})$ is the logistic sigmoid activation function, and $I$ represents a constant external input. It should be noted that although it is standard to include these external input terms, we consider CTRNNs with them omitted following \cite{funahashi}. 

By Theorems 1-3 in \cite{funahashi}, Einstein's equations (\ref{eq:evo1})-(\ref{eq:evo4}) take their neural network form as:
\begin{eqnarray}
\label{eq:ctrnn01}
\Sigma_{+}' &=& \frac{1}{\tau_1} \left[-\Sigma_{+} + w_{1}^{1} \sigma(\Sigma_{+} + \theta_1)  + w_{1}^{2} \sigma(\Sigma_{-} + \theta_2) + w_{1}^{3} \sigma(N_1 + \theta_3) + w_{1}^{4}\sigma(N_2 + \theta_4) + w_{1}^{5}\sigma(N_3 + \theta_5)\right], \\
\label{eq:ctrnn2}
\Sigma_{-}' &=& \frac{1}{\tau_2} \left[-\Sigma_{-} + w_{2}^{1} \sigma(\Sigma_{+} + \theta_1)  + w_{2}^{2} \sigma(\Sigma_{-} + \theta_2) + w_{2}^{3} \sigma(N_1 + \theta_3) + w_{2}^{4}\sigma(N_2 + \theta_4) + w_{2}^{5}\sigma(N_3 + \theta_5)\right], \\
\label{eq:ctrnn3}
N_1' &=& \frac{1}{\tau_3} \left[-N_1 + w_{3}^{1} \sigma(\Sigma_{+} + \theta_1)  + w_{3}^{2} \sigma(\Sigma_{-} + \theta_2) + w_{3}^{3} \sigma(N_1 + \theta_3) + w_{3}^{4}\sigma(N_2 + \theta_4) + w_{3}^{5}\sigma(N_3 + \theta_5)\right], \\
\label{eq:ctrnn4}
N_2' &=& \frac{1}{\tau_4} \left[-N_2+ w_{4}^{1} \sigma(\Sigma_{+} + \theta_1)  + w_{4}^{2} \sigma(\Sigma_{-} + \theta_2) + w_{4}^{3} \sigma(N_1 + \theta_3) + w_{4}^{4}\sigma(N_2 + \theta_4) + w_{4}^{5}\sigma(N_3 + \theta_5)\right], \\
\label{eq:ctrnn5}
N_3' &=& \frac{1}{\tau_5} \left[-N_3 + w_{5}^{1} \sigma(\Sigma_{+} + \theta_1)  + w_{5}^{2} \sigma(\Sigma_{-} + \theta_2) + w_{5}^{3} \sigma(N_1 + \theta_3) + w_{5}^{4}\sigma(N_2 + \theta_4) + w_{5}^{5}\sigma(N_3 + \theta_5)\right].
\end{eqnarray}
What is interesting about Eqs. (\ref{eq:ctrnn01}) - (\ref{eq:ctrnn5}) is that the complicated nonlinear Einstein equations have now been written as relatively simpler equations involving simple linear combinations of the weights $w^j_i$, logistic sigmoid activation functions which are bounded between 0 and 1, positive constants $\tau_i$, and bias terms $\theta_i$. The task at hand is to find values for these parameters that allow us to match the right-hand-sides of Eqs. (\ref{eq:ctrnn01}) - (\ref{eq:ctrnn5}) to the right-hand-sides of Eqs. (\ref{eq:evo1})-(\ref{eq:evo4}). 

We use the following methodology to find optimal values for these parameters. We wrote a numerical algorithm in MATLAB that uses the \emph{ode45} solver, which is an explicit Runge-Kutta solver to sample several solution curves generated from Eqs. (\ref{eq:evo1})-(\ref{eq:evo4}). We then used a nonlinear least-squares method to find optimal parameters that minimized the Euclidean distance between the right-hand-sides of Eqs. (\ref{eq:evo1})-(\ref{eq:evo4}) and  Eqs. (\ref{eq:ctrnn01}) - (\ref{eq:ctrnn5}). 


In what follows, we will examine the solutions to Eqs. (\ref{eq:ctrnn01}) - (\ref{eq:ctrnn5}) for some Bianchi class A models by performing numerical experiments and determining if the neural Einstein equations exhibit some well-known properties in terms of their dynamical behaviour and asymptotic states with respect to isotropization.

\subsection{Bianchi Type I Models}
Bianchi Type I models are the anisotropic generalization of spatially flat Friedmann-Lemaitre-Robertson-Walker (FLRW) models. They are defined by $N_1 = N_2 = N_3 = 0$. From Eqs. (\ref{eq:ctrnn01}) - (\ref{eq:ctrnn5}), the dynamics of such a model is given by:
\begin{eqnarray}
\label{eq:bianchi11}
\Sigma_{+}' &=& \frac{1}{\tau_1} \left[-\Sigma_{+} + w_{1}^{1}\sigma(\Sigma_{+} + \theta_1) + w_{1}^{2} \sigma(\Sigma_{-} + \theta_2)\right], \\
\label{eq:bianchi12}
\Sigma_{-}' &=& \frac{1}{\tau_2} \left[-\Sigma_{-} + w_{2}^{1} \sigma(\Sigma_{+} + \theta_1)  + w_{2}^{2} \sigma(\Sigma_{-} + \theta_2)\right].
\end{eqnarray}

Choosing $\gamma = 4/3$ corresponding to radiation-dominated matter and following the aforementioned minimization procedure, we obtain the following parameters in Eqs. (\ref{eq:bianchi11}) - (\ref{eq:bianchi12}):
\begin{equation}
\label{eq:results1}
\tau_1 = 1.445, \quad w_1^1 = 0.028, \quad \theta_1 = -31.598, \quad w_2^1 = -0.0073, \quad \theta_2 = 0.2842,  \quad \tau_2 = 1.444, \quad w_2^2 = -0.041, \quad w_1^2 = 0.0098.
\end{equation}
Solution curves corresponding to Eqs. (\ref{eq:bianchi11})-(\ref{eq:results1}) are shown in Fig. \ref{Fig:fig1} below.
\begin{figure}[h]
  \centering
    \includegraphics[scale=0.60]{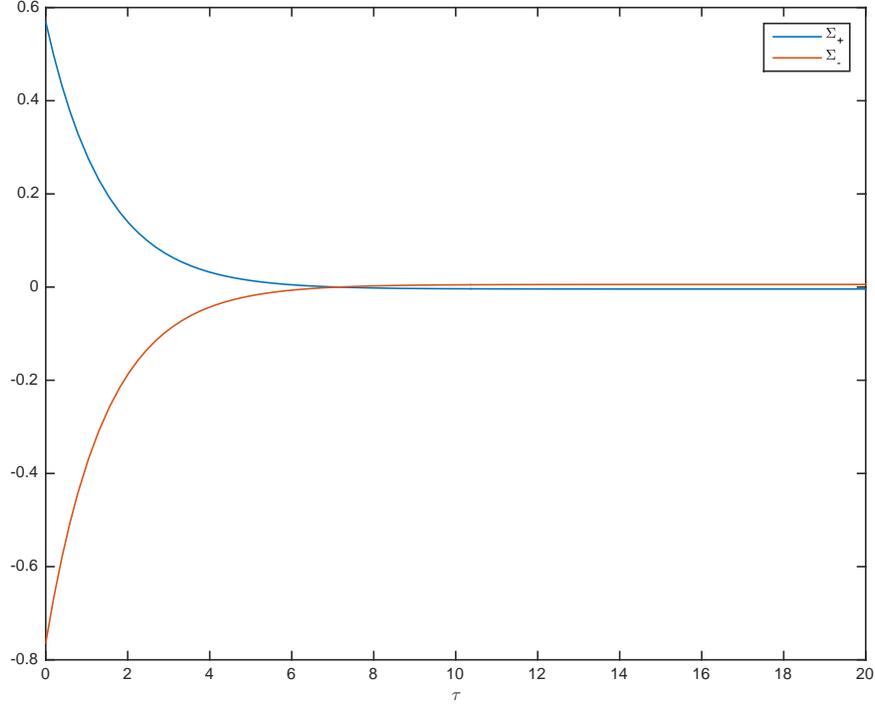}
    \caption{Some solution curves to Eqs. (\ref{eq:bianchi11})-(\ref{eq:results1}) for $\gamma = 4/3$.}
     \label{Fig:fig1}
\end{figure}
In Fig. \ref{Fig:fig1}, we show some solutions of the Bianchi Type I CTRNN equations and that they exhibit the expected dynamical behaviour. In particular, the solution curves are approaching $(\Sigma_{+}, \Sigma_{-}) = (0,0)$ which is the flat FLRW universe. 
\newpage

Choosing $\gamma = 0$ corresponding to a cosmological constant, and following the aforementioned minimization procedure, we obtain the following parameters in Eqs. (\ref{eq:bianchi11}) - (\ref{eq:bianchi12}):
\begin{equation}
\label{eq:results2}
\tau_1 = 0.4725, \quad w_1^1 = 1.4712, \quad \theta_1 = -5.318, \quad w_2^1 = -0.0092, \quad \theta_2 = 11.5481,  \quad \tau_2 = 0.4726, \quad w_2^2 = -1.9998, \quad w_1^2 = 0.0125.
\end{equation}
Solution curves corresponding to Eqs. (\ref{eq:bianchi11})- (\ref{eq:bianchi12}), (\ref{eq:results2}) are shown in Fig. \ref{Fig:fig2} below.
\begin{figure}[h]
  \centering
    \includegraphics[scale=0.60]{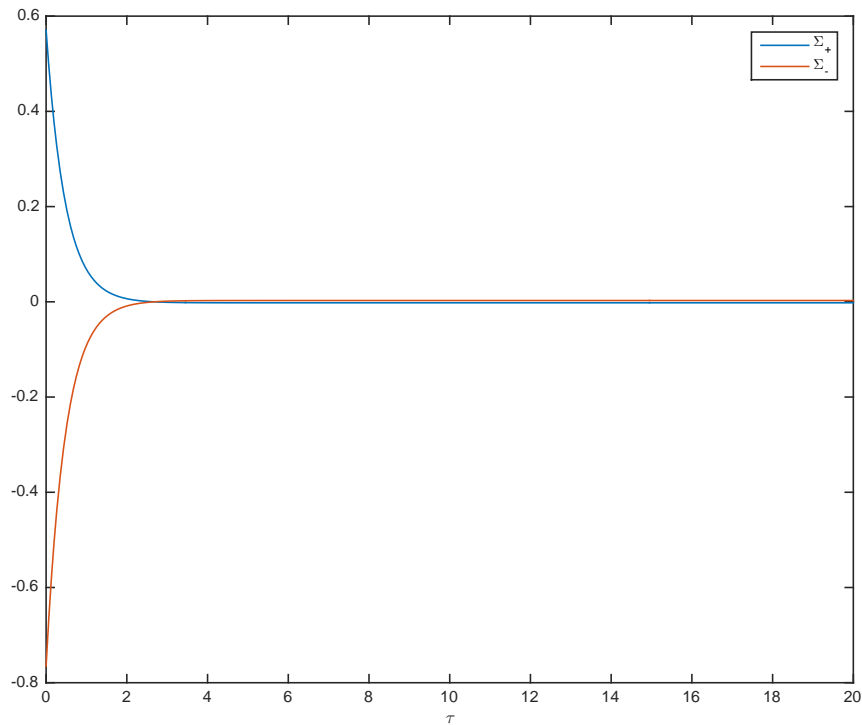}
    \caption{Some solution curves corresponding to Eqs. (\ref{eq:bianchi11})- (\ref{eq:bianchi12}), (\ref{eq:results2}) for $\gamma = 0$, a cosmological constant.}
     \label{Fig:fig2}
\end{figure}
In Fig. \ref{Fig:fig2}, we show some solutions of the Bianchi Type I CTRNN equations and that they exhibit the expected dynamical behaviour. In particular, the solution curves are approaching the point $(\Sigma_{+}, \Sigma_{-}) = (0,0)$ which is the flat FLRW universe. 
\newpage

We finally consider dust models $\gamma = 1$. We obtain the following parameters in Eqs. (\ref{eq:bianchi11}) - (\ref{eq:bianchi12}):
\begin{equation}
\label{eq:results3}
\tau_1 = 0.9572, \quad w_1^1 = 0.0061, \quad \theta_1 = -32.89, \quad w_2^1 = -0.1591, \quad \theta_2 = -3.8632,  \quad \tau_2 = 0.9567, \quad w_2^2 = -0.0025, \quad w_1^2 = 0.2123.
\end{equation}
Solution curves corresponding to Eqs. (\ref{eq:bianchi11})- (\ref{eq:bianchi12}), (\ref{eq:results3}) are shown in Fig. \ref{Fig:fig3} below.
\begin{figure}[h]
  \centering
    \includegraphics[scale=0.60]{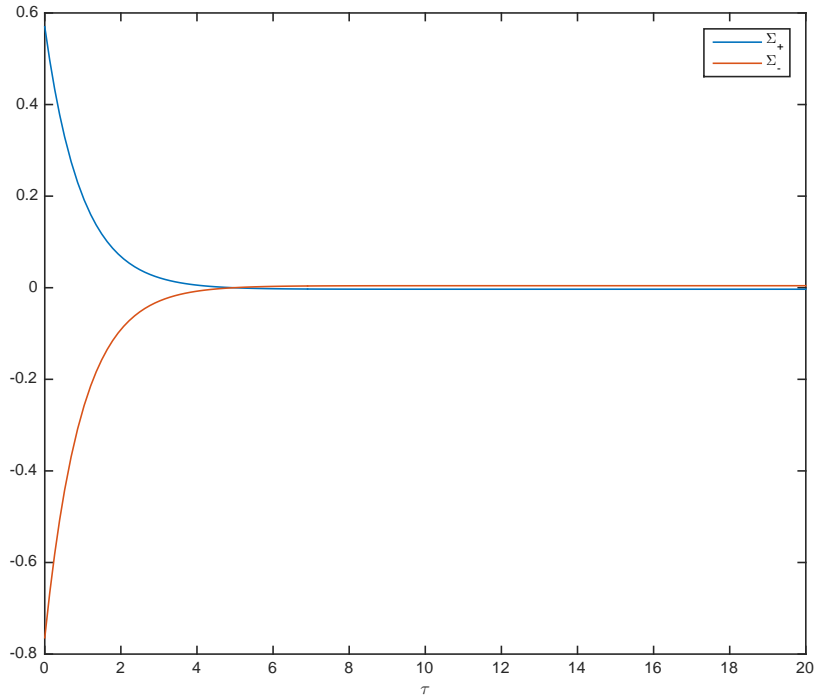}
    \caption{Some solution curves corresponding to Eqs. (\ref{eq:bianchi11})- (\ref{eq:bianchi12}), (\ref{eq:results3}) for $\gamma = 1$, pressure-free dust.}
     \label{Fig:fig3}
\end{figure}
One sees from Fig. \ref{Fig:fig3} that the Bianchi Type I CTRNN equations exhibit the expected dynamical behaviour. In particular, the solution curves are approaching the point $(\Sigma_{+}, \Sigma_{-}) = (0,0)$ which is the flat FLRW universe. 
\newpage

\subsection{Bianchi Type II Models}
We now examine Bianchi Type II models, which are defined by $N_1 > 0, N_2 = N_3 = 0$. 
From Eqs. (\ref{eq:ctrnn01}) - (\ref{eq:ctrnn5}), the dynamics of such a model is given by:
\begin{eqnarray}
\label{eq:ctrnn21}
\Sigma_{+}' &=& \frac{1}{\tau_1} \left[-\Sigma_{+} + w_{1}^{1} \sigma(\Sigma_{+} + \theta_1)  + w_{1}^{2} \sigma(\Sigma_{-} + \theta_2) + w_{1}^{3} \sigma(N_1 + \theta_3) \right], \\
\label{eq:ctrnn22}
\Sigma_{-}' &=& \frac{1}{\tau_2} \left[-\Sigma_{-} + w_{2}^{1} \sigma(\Sigma_{+} + \theta_1)  + w_{2}^{2} \sigma(\Sigma_{-} + \theta_2) + w_{2}^{3} \sigma(N_1 + \theta_3)\right], \\
\label{eq:ctrnn23}
N_1' &=& \frac{1}{\tau_3} \left[-N_1 + w_{3}^{1} \sigma(\Sigma_{+} + \theta_1)  + w_{3}^{2} \sigma(\Sigma_{-} + \theta_2) + w_{3}^{3} \sigma(N_1 + \theta_3) \right].
\end{eqnarray}


Choosing $\gamma = 4/3$ corresponding to radiation-dominated matter, following the aforementioned minimization procedure, we obtain the following parameters in Eqs. (\ref{eq:ctrnn21}) - (\ref{eq:ctrnn23}):
\begin{eqnarray}
\label{eq:results4}
\tau_1 = 0.6152, \quad w_1^1 = -0.5688 \quad \theta_1 = 28.901, \\ \nonumber
\quad w_2^1 = 1.9849, \quad \theta_2 = -0.2793, \quad w_3^1 = -0.0334, \\ \nonumber
 \quad \theta_3 = -11.9495, \quad \tau_2 = 1.3591, \quad w_1^2 = -0.0913, \\  \nonumber
 \quad w_2^2 = 0.0395, \quad w_3^2 = 0.088, \quad \tau_3 = 0.1587, \quad w_1^3 = 0.0978, \quad w_2^3 = 2, \quad w_3^3 = 0.5312. \nonumber
\end{eqnarray}
Solution curves corresponding to  Eqs. (\ref{eq:ctrnn21}) - (\ref{eq:ctrnn23}), (\ref{eq:results4}) are shown in Fig. \ref{Fig:fig4} below.
\begin{figure}[h]
  \centering
    \includegraphics[scale=0.60]{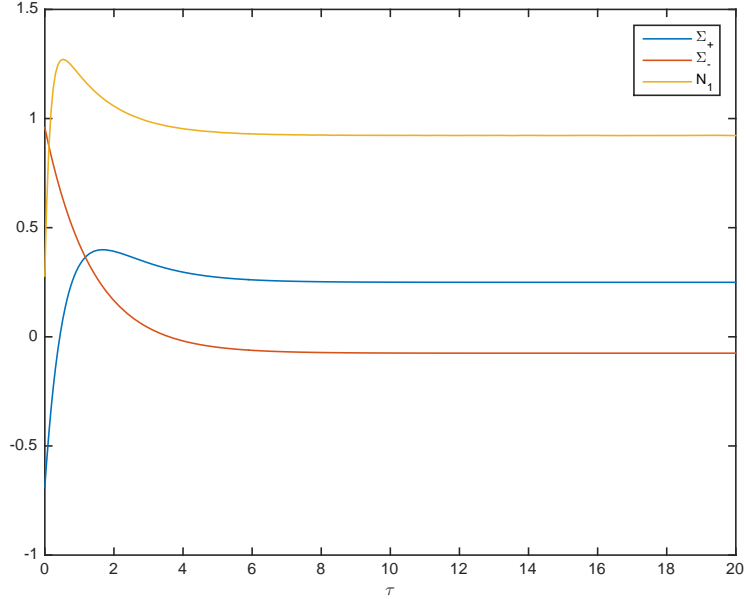}
    \caption{Some solution curves corresponding to Eqs. (\ref{eq:ctrnn21}) - (\ref{eq:ctrnn23}), (\ref{eq:results4}) for $\gamma = 4/3$, radiation-dominated matter.}
     \label{Fig:fig4}
\end{figure}
One sees from Fig. \ref{Fig:fig4} that the Bianchi Type II CTRNN equations exhibit the expected dynamical behaviour. In particular, the model does \emph{not} isotropize, and goes towards a solution where $\Sigma_- \to 0$, $N_1 >0$, and $\Sigma_{+} \neq 0$. This behaviour is what is expected from such models which respect to future-time evolution as discussed in \cite{ellis}.
\newpage


Choosing $\gamma = 1$ corresponding to pressure-free dust, we obtain the following parameters in Eqs. (\ref{eq:ctrnn21}) - (\ref{eq:ctrnn23}):
\begin{eqnarray}
\label{eq:results6}
\tau_1 = 0.5809, \quad w_1^1 = 0.0581 \quad \theta_1 = -40.911, \\ \nonumber
\quad w_2^1 = 1.9973, \quad \theta_2 = -0.538, \quad w_3^1 = -0.735, \\ \nonumber
 \quad \theta_3 = 1.571, \quad \tau_2 = 0.802, \quad w_1^2 = 0.0735, \\  \nonumber
 \quad w_2^2 = 2, \quad w_3^2 = -0.992, \quad \tau_3 = 0.1762, \quad w_1^3 = 0.491, \quad w_2^3 = 1.999, \quad w_3^3 = 0.1302. \nonumber
\end{eqnarray}
Solution curves corresponding to Eqs. (\ref{eq:ctrnn21}) - (\ref{eq:ctrnn23}), (\ref{eq:results6}) are shown in Fig. \ref{Fig:fig6} below.
\begin{figure}[h]
  \centering
    \includegraphics[scale=0.40]{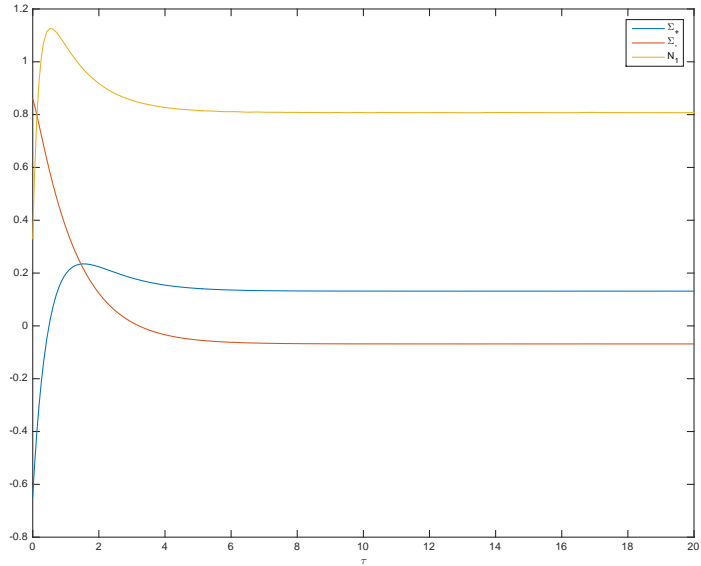}
    \caption{Some solution curves corresponding to Eqs. (\ref{eq:ctrnn21}) - (\ref{eq:ctrnn23}), (\ref{eq:results6}) for $\gamma = 1$, a cosmological constant.}
     \label{Fig:fig6}
\end{figure}
One sees from Fig. \ref{Fig:fig6} that the Bianchi Type II CTRNN equations exhibit the expected dynamical behaviour. In particular, the model does \emph{not} isotropize, and goes towards a solution where $\Sigma_- \to 0$, $N_1 >0$, and $\Sigma_{+} \neq 0$. This behaviour is what is expected from such models which respect to future-time evolution as discussed in \cite{ellis}.
\newpage

\section{Discussion of Results}
\begin{figure}[h]
  \centering
    \includegraphics[scale=0.40]{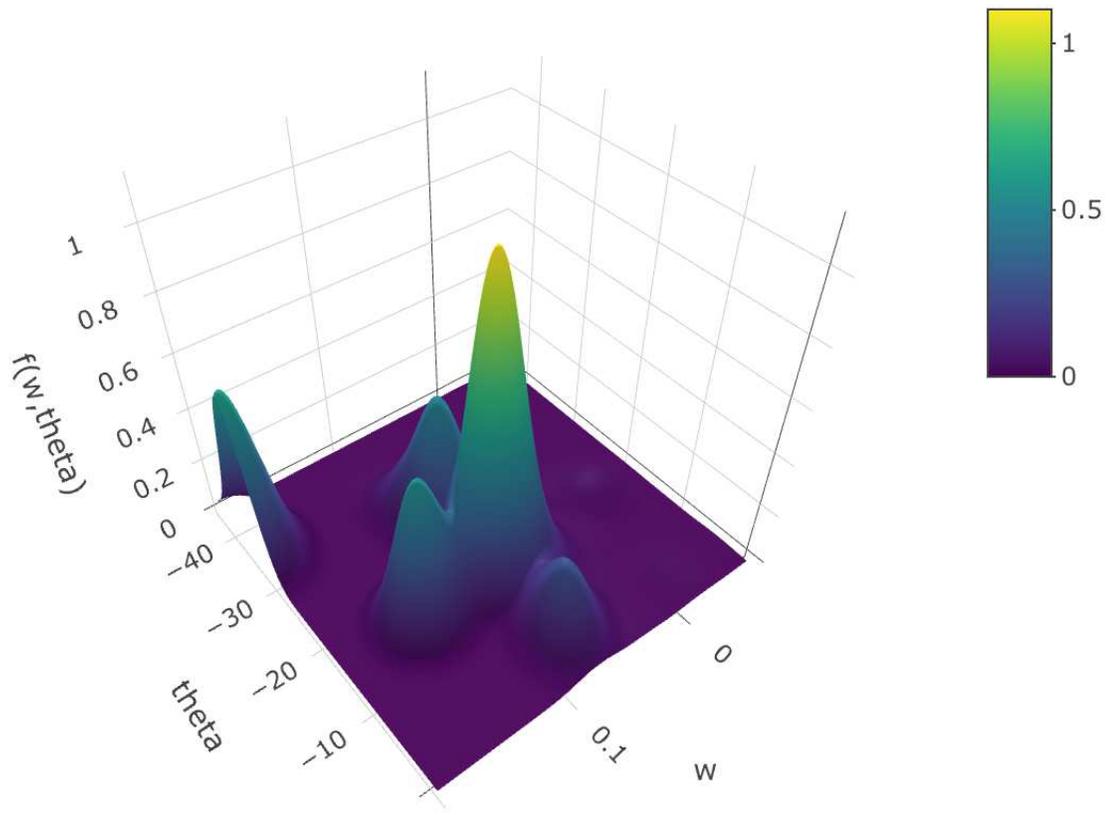}
    \caption{A plot showing the bivariate probability distribution $f(w,\theta)$ resulting from fitting Eq. (\ref{eq:ctrnnFL}) to Eq. (\ref{eq:FLdyn1}) for several initial conditions corresponding to $\gamma = 4/3$.}
     \label{Fig:fig7}
\end{figure}

The results so far have shown that CTRNNs have an ability to learn the dynamical behaviour of Einstein's equations. However, as the reader may suspect, doing so depends highly on the initial conditions supplied to Eqs. (\ref{eq:evo1})-(\ref{eq:evo4}), in addition to the equation of state parameter $\gamma$. To address this issue, we look at an example where we apply a CTRNN to the first-order Einstein equation for FLRW models \cite{ellis}:
\begin{equation}
\label{eq:FLdyn1}
\Omega' = -(3\gamma - 2)(1-\Omega)\Omega,
\end{equation}
where $\Omega$ is the density parameter. Namely, $\Omega > 1$ corresponds to closed $(k = +1)$ models, $\Omega = 1$ corresponds to flat $(k = 0)$ models, and $\Omega < 1$ corresponds to open $(k = -1)$ models. From Eq. \ref({eq:ctrnn1}), the corresponding to CTRNN is:
\begin{equation}
\label{eq:ctrnnFL}
\Omega' = \frac{1}{\tau}\left[-\Omega + w \sigma\left(\Omega + \theta\right) \right],
\end{equation}
where $w$ and $\theta$ are the weight and bias parameters to be learned. We replicate the aforementioned optimization procedure but in our numerical algorithm we simulated solutions to these equations for 2000 initial conditions. What one gets then is a multivariate distribution for the space of optimal parameters. We use kernel density estimation to estimate this density function $f(w,\theta)$ in Fig. (\ref{Fig:fig7}).
\newpage
One therefore sees that there is a space of values for the neural network weights that can reproduce the dynamics of Einstein's equations for a very large number of initial conditions. This suggests that in the full picture described by  Eqs. (\ref{eq:ctrnn01}) - (\ref{eq:ctrnn5}), there is a probability density function of 55 parameters that when sampled and substituted into these equations
fully replicates the dynamics of the Einstein equations (\ref{eq:evo1})-(\ref{eq:evo4}) for any number of initial conditions. 

\section{Conclusions}
In this paper, we have discussed how one can write Einstein's field equations for spatially homogeneous cosmological models as continuous-time recurrent neural networks (CTRNNs), which are simple linear combinations of a weight values and logistic sigmoid functions. In doing so, we are able to write complicated nonlinear dynamical systems in a simpler form leading to a potentially easier way to analyze the dynamical behaviour of cosmological solutions to Einstein's equations.

The work presented here is a starting point and is by no means complete. In future work, we plan to extend the ideas here to Bianchi Class B cosmologies and the $G_2$ class of cosmologies which are systems of partial differential equations. Using a CTRNN to write Einstein's equations for the latter would allow us to once again write a system of partial differential equations as a system of simpler ordinary differential equations.

\newpage 
\bibliography{sources}

\end{document}